\documentclass[aps,prl,preprint,amsmath,amsfonts,amssymb, braket, groupedaddress,floatfix]{revtex4-2}

\usepackage{graphicx}
\usepackage{dcolumn}
\usepackage{bm}
\usepackage{hyperref}
\usepackage[usenames,dvipsnames]{color}
\usepackage[normalem]{ulem}
\usepackage{amsmath}
\usepackage{amssymb}
\usepackage{braket}
\usepackage{slashed}

\begin{document}


\title{Can a Spin Liquid State Persist as the Ground State in the Presence of Competing Interactions and Disorder?}

\author{Sumanta Mukherjee$^{1,*}$}

\affiliation{$^1$ Solid State and Structural Chemistry Unit, Indian Institute of Science, Bengaluru, Karnataka 560012, India\\}

\email{sm31081985@gmail.com}

\date{\today}

\begin{abstract}
In this report, we have shown that, in an otherwise geometrically frustrated lattice, the underlying interactions compete to stabilize different magnetic phases. In conjunction with fluctuations, these interactions may lead to the formation of unusual ordered phases, providing a pathway for understanding the fluctuation-driven order-by-disorder phenomenon. This competition is further modified by the presence of inherent structural disorder in a frustrated two-dimensional lattice. Furthermore, due to the local nature of the additional structural disorder, the majority of the samples evolve toward glassy dynamics, which can not only mimic spin-liquid-like behavior but also make it difficult to identify genuine spin-liquid candidates experimentally.
\end{abstract}

\maketitle

\section{Introduction}

Geometric frustration\cite{ref-1,ref-2} in a magnetic lattice can give rise to a variety of exotic phenomena\cite{ref-1,ref-2,ref-3}, including unconventional spin-glass\cite{ref-4,ref-5,ref-6,ref-7} and spin-liquid phases\cite{ref-8,ref-9,ref-10,ref-11,ref-12}. A simple example of geometric frustration is provided by a triangular lattice with nearest-neighbor antiferromagnetic interactions\cite{ref-8,ref-9,ref-11,ref-12}. Ever since the proposal of the resonating valence bond (RVB) state\cite{ref-1,ref-13,ref-14} in such systems, there has been sustained interest in the search for candidate materials over the past few decades. The interest in these systems became more pronounced following the suggestion that a clean (structural-disorder-free) frustrated lattice may not exhibit any magnetic ordering or spin freezing down to the lowest accessible temperatures, thereby remaining in a liquid-like state even at temperatures approaching absolute zero\cite{ref-1,ref-2,ref-15,ref-16}. Importantly, in the absence of long-range magnetic order, it has been proposed that such materials do not support conventional Goldstone spin-wave excitations but are instead characterized by fractionalized spinon excitations\cite{ref-1,ref-2,ref-17}, which may possess a metal-like Fermi surface\cite{ref-16,ref-18}. The primary focus of spin-liquid research is the identification of resonating valence bond (RVB) states, valence-bond (VB) glass\cite{ref-7}, and fractionalized spinon excitations\cite{ref-16,ref-18}, whether gapped or gapless, using a variety of experimental techniques\cite{ref-8,ref-10,ref-11,ref-15,ref-17}. The large ground-state degeneracy associated with these fluid-like states provides a possible experimental basis for identifying such systems\cite{ref-16,ref-18}. Although geometrically frustrated magnetic systems are abundant, a few representative examples include Volborthite\cite{ref-5}, k-(BEDT-TTF)$_2$Cu$_2$(CN)$_3$\cite{ref-8}, EtMe$_3$Sb[Pd(dmit)$_2$]$_2$\cite{ref-9}, and herbertsmithite\cite{ref-10}.\\
While the identification and characterization of spin-liquid materials remain active areas of research, several important questions arising from their behavior have become topics of considerable interest and debate in recent years. One of the foremost questions concerns the role of structural disorder, which is inherently present in all synthesized materials\cite{ref-5,ref-19,ref-20,ref-21,ref-22}. While it is now reasonably well understood from both experimental\cite{ref-23} and theoretical studies\cite{ref-19,ref-21} that the presence of structural disorder can induce glassy behavior in an otherwise spin-liquid candidate, it has also been suggested that, under certain circumstances, structural disorder can destabilize the spin-liquid state and instead promote magnetic ordering that is absent in the pristine frustrated lattice\cite{ref-24,ref-25}. This observation points to an order-by-structural-disorder mechanism, a phenomenon that has been the subject of extensive theoretical and experimental investigation over the past several decades\cite{ref-24,ref-25,ref-26,ref-27,ref-28}. However, only a few examples of order-by-structural-disorder phenomena have been reported in this context\cite{ref-29}, with the majority of systems instead exhibiting glassy behavior\cite{ref-23}. A second important question concerns the role of fluctuations in determining the properties of these materials. Although this aspect has not yet been explored extensively, it is generally believed\cite{ref-30} that geometric frustration can enhances the role of fluctuations, thereby suppressing conventional magnetic ordering tendencies.\\
In this report, we investigate these aspects from a field-theoretical perspective. We show that, in a generic spin-liquid candidate, the competing interactions and underlying symmetries give rise to several distinct magnetic structures (hereafter referred to as magnetic phases), which differ in the geometry (symmetry) of their magnetic order parameters. As geometric frustration destabilizes the conventional ground-state configuration, these competing magnetic phases become nearly degenerate and compete to establish the stable ground state. While these phases may not possess stable minima away from the order-parameter-zero state, the induced renormalization\cite{ref-31} of the Landau coefficients\cite{ref-32,ref-33,ref-34} can shift the free-energy landscape and generate minima at finite values of the order parameter. Therefore, magnetic ordering may emerge in the system even though none of the initial phases possesses a stable minimum at a finite value of the order parameter. This mechanism shares similarities with the Coleman-Weinberg mechanism\cite{ref-35}, where fluctuations modify the effective free-energy landscape and generate symmetry-breaking minima. This result provides a field-theoretical framework for understanding fluctuation-induced order-by-disorder phenomena in geometrically frustrated systems.\\
However, since inherent structural disorder (e.g., defects) can influence the interactions among distinct phases by locally modifying their competition, the order parameters of these phases acquire spatial randomness. This necessitates the use of a local loop correction\cite{ref-31,ref-36} in the renormalization procedure, which captures the resulting inhomogeneities in the competing fields. A glassy state naturally emerges from the disorder-induced inhomogeneities. Using a simplified replica model\cite{ref-19,ref-37}, we find that, in the dilute limit and at very low temperatures, the glass transition temperature scales with the interaction strength and the disorder variance parameter induced by the loop correction. We verify this outcome through comparison with known experimental results. The presented scenario therefore provides a possible explanation for the emergence of glassy dynamics in many spin-liquid candidates at very low temperatures. However, since the proposed model relies entirely on ordering tendencies induced by fluctuations \cite{ref-31}, its validity remains restricted to the low-disorder and low-temperature limits.

\section{Results and Discussion}

To begin the discussion, we consider a prototypical example of a geometrically frustrated system (\textbf{Figure 1a}). Although many well-known candidates exhibit ordering below a certain temperature, we choose a two-dimensional triangular system as a model system to maintain generality. We impose the condition that the system remains in a liquid state down to the lowest temperatures, and can therefore be treated as a fluid throughout the accessible temperature range. A simplistic free energy form representing a liquid state\cite{ref-32,ref-33,ref-38} can be written as
\begin{equation}
F = a\phi_l^4 + b\phi_l^2
\end{equation}
where ($\phi_l$) is an order parameter related to magnetization, with the expectation value ($\langle \phi_l \rangle = 0$), and ($b > 0$) at all temperatures. In a general description, the free energy expression for an order parameter incorporates the relevant symmetries of the order parameter. However, in the present case, we are modeling an order parameter associated with an unknown magnetic texture (i.e., an unknown Fourier expansion of the magnetization structure), which predominantly remains in a liquid state. Therefore, we begin with this simplified free energy expression and introduce additional complexities in subsequent derivations as required. Before incorporating the effects of other phases into the free-energy expression, we first examine how a system can generally exhibit different phases \cite{ref-32} and how their competition may influence the free-energy landscape.
As a crude analogy, consider a system that undergoes a transition from a liquid phase and develops an orthorhombic structure under conventional cooling conditions. However, under a rapid cooling process, the same system may instead retain a cubic structure. Within the conventional theoretical framework, a liquid state is considered an isotropic and highly symmetric phase of a system. However, from a density perspective, the liquid density\cite{ref-32} can be expressed as $\rho(\mathbf{r}) = \rho_0 + \delta\rho(\mathbf{r})$, where the density fluctuation ($\delta\rho(\mathbf{r})$) can be expanded in terms of its Fourier components. In the presence of competing multiphase order parameters, these Fourier components can be reorganized as a sum over the Fourier modes associated with each type of order parameter field\cite{ref-32}. Consequently, the free energy expression can be written in the following form\cite{ref-39,ref-40}.
The free energy can be expressed as
\begin{equation}
F = F[\phi_1] + F[\phi_2] + g f(\phi_1,\phi_2),
\end{equation}
where the possible expectation values of the order parameters, ($\langle \phi_1 \rangle,\langle \phi_2 \rangle$) $= (0,0), (m_a,0)$, $\text{ and } (0,m_b)$, correspond to the liquid state, the ($\phi_1$)-ordered state, and the ($\phi_2$)-ordered state, respectively\cite{ref-39}. The sign and magnitude of the coupling parameter ($g$) in the interaction term determine whether the two phases compete with or coexist alongside each other\cite{ref-39}. The competing phases need not be limited to two order parameters, as considered above. More generally, one can introduce multiple order parameters and express the free energy as
\begin{figure}[t]
\begin{center}
\includegraphics[width=1.0\columnwidth]{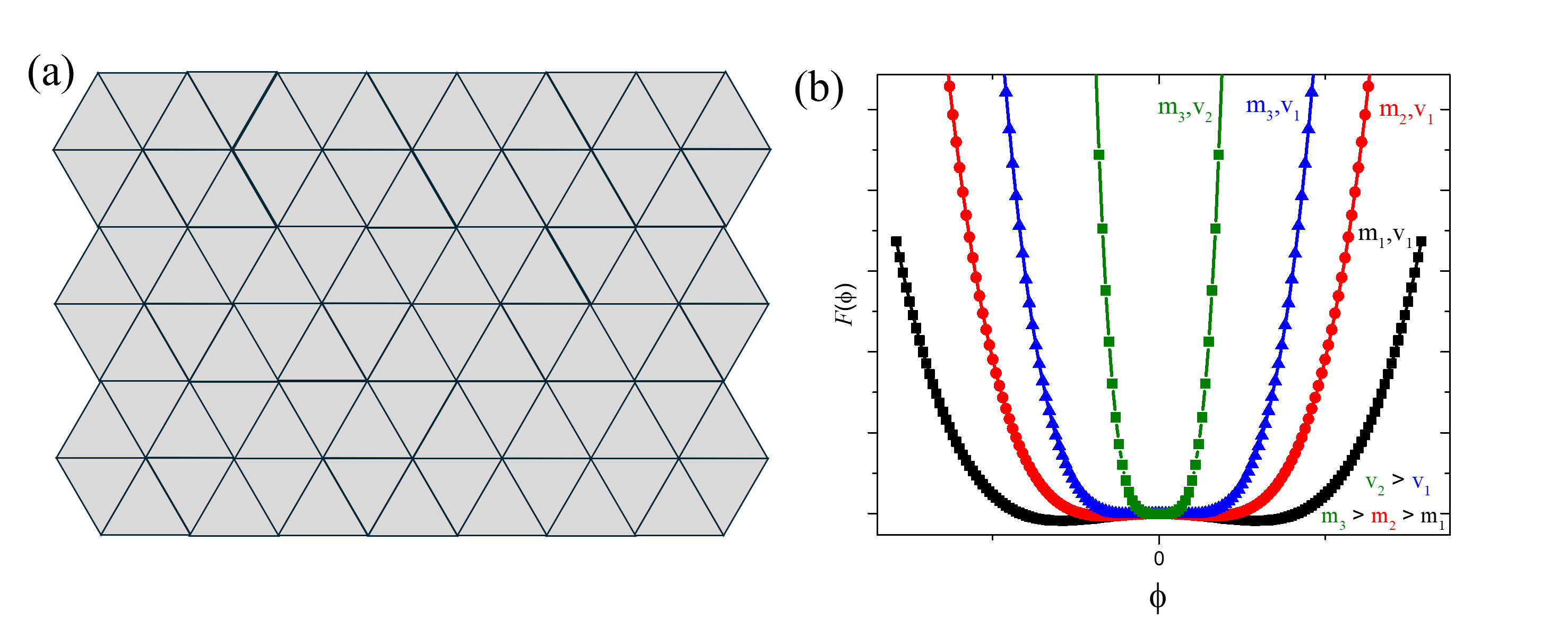}
\caption{(a) Schematic representation of the triangular lattice. (b) Systematic shift of the free-energy minimum from ($\phi = 0$) to ($\phi = \phi_0$), as derived from \textbf{Equation 13}, for representative values of the parameters $v$, $g$, $w$, $m$ and $c$.}
\label{fig1}
\end{center}
\end{figure}
\begin{equation}
F = \sum_i F_i[\phi_i] + \sum_{i\neq j} g_{ij} f(\phi_i,\phi_j)
\end{equation}
While minimizing the free energy in the presence of all possible phases is a challenging task, we restrict our phenomenological model to remain tractable by considering only the phases with the dominant contributions to the free energy. At this stage, we neglect the detailed symmetries and spatial inhomogeneities associated with these phases. However, we note that incorporating the symmetries of the order parameters is essential for a detailed description of the system, as these symmetries determine why specific ordering patterns, such as 120-degree coplanar antiferromagnetic order\cite{ref-41}, ferro- or antiferro-quadrupolar order\cite{ref-41}, cycloidal order\cite{ref-42,ref-43}, or spiral order\cite{ref-44,ref-45}, are stabilized. In the present work, we maintain a phenomenological approach and defer a more detailed treatment of these symmetry-related aspects to future studies. Within this framework, the total free energy can be written as
\begin{equation}
F[\phi_1,\phi_2] = F[\phi_1] + F[\phi_2] + g f(\phi_1,\phi_2)
\end{equation}
or, explicitly,
\begin{equation}
F[\phi,\psi] = \frac{c}{2}|\nabla\psi|^2 + \frac{m}{2}\psi^2 + \frac{u}{24}\psi^4 + \frac{v}{24}\phi^4 + \frac{w}{2}\phi^2 + g\phi^2\psi^2
\end{equation}
Where $c$, $m$, $u$, $v$ and $w$ are coefficients. Note that the last term represents the interaction between the order parameters, which is constructed based on the overall symmetry of the coupled fields. We further impose the condition that the expectation value of the order parameter satisfies ($\langle \psi \rangle = 0$) and that ($m > 0$) at all temperatures. Therefore, this field does not undergo an ordering transition, like the other phase.
To extend the standard renormalization approach\cite{ref-31,ref-35,ref-38}, exemplified by the Coleman-Weinberg formalism\cite{ref-35}, to the Ginzburg-Landau framework\cite{ref-33,ref-34}, we employ the well-known procedure of expressing the Ginzburg-Landau free-energy functional as a field-theoretic partition function. This extension enables the incorporation of fluctuation corrections beyond the mean-field approximation and facilitates the evaluation of the renormalized effective potential associated with the order-parameter field. The partition function takes the general form:
\begin{equation}
Z = \int \mathcal{D}[\psi]\mathcal{D}[\phi] \exp\left(-\beta F[\phi,\psi]\right).
\end{equation}
Where $\beta = 1/k_BT$, and $k_B$ being the Boltzmann constant. To account for the effects of fluctuations around the mean-field solution, we expand the ($\psi$) field around its classical configuration by writing $\psi = \psi_c + \zeta$, where ($\zeta$) represents the fluctuations around the mean-field configuration ($\psi_c$). The free energy can then be expanded as $F[\psi] = F[\psi_c] + \frac{1}{2}\zeta M \zeta + \cdots$, where ($M$) is the Hessian matrix, which is related to the second functional derivative of the free energy. By integrating over the fluctuations using the identity
\begin{equation}
\int \mathcal{D}[\zeta] \exp\left(-\frac{\beta}{2}\zeta M\zeta\right) = (\det \beta M)^{-1/2}
\end{equation}
with
\begin{equation}
M = -c\nabla^2 + m + \frac{u}{2}\psi_c^2 + 2g\phi^2,
\end{equation}
we obtain the effective free energy:
\begin{equation}
F_{\mathrm{eff}} = F[\phi,\psi_c] + \frac{1}{2\beta}\mathrm{Tr}\ln(\beta M)
\end{equation}
Expanding around the symmetric phase ($\psi_c=0$) gives the effective free energy as
\begin{equation}
F_{\mathrm{eff}}(\phi)=\frac{v}{24}\phi^4 + \frac{w}{2}\phi^2+\frac{1}{2\beta}\mathrm{Tr}\ln\left\{\beta(-c\nabla^2 + m + 2g\phi^2)\right\}
\end{equation}
Evaluating the  ($\mathrm{Tr}\ln$) term in (2+1)-dimensional Euclidean space leads to the following expression for the effective free energy:
\begin{equation}
F_{\mathrm{eff}}(\phi)=\frac{v}{24}\phi^4+\frac{w}{2}\phi^2-\frac{1}{12\pi c^{3/2}}(m+2g\phi^2)^{3/2}
\end{equation}
Therefore, within this simplified analogy, we may conclude that fluctuations in the ($\psi$) field introduce an additional effective potential contribution into the free-energy expression of the ($\phi$) field.
We may further expand the potential term in the following limit, ($m \gg g\phi^2$), as
\begin{equation}
(m+2g\phi^2)^{3/2}\approx m^{3/2}+3gm^{1/2}\phi^2+\frac{3g^2}{2m^{1/2}}\phi^4+\cdots
\end{equation}
Substituting this expansion into the effective free-energy expression yields a renormalized Landau free energy with modified coefficients for the quadratic and quartic terms.
\begin{equation}
F_{\mathrm{eff}}(\phi)=\left(\frac{v}{24}-\frac{g^2}{8\pi c^{3/2}m^{1/2}}\right)\phi^4+
\left(\frac{w}{2}-\frac{gm^{1/2}}{4\pi c^{3/2}}\right)\phi^2 .
\end{equation}
It is evident from this expression that, although the initial free-energy expression possesses a minimum only at ($\phi=0$), the loop corrections arising from fluctuations may generate a stable minimum at ($\phi=\phi_0$). This provides the possibility of a fluctuation-induced symmetry breaking in the system. We provide a pictorial representation of the emergence of a stable minimum away from ($\phi=0$) in \textbf{Figure 1(b)} for a representative choice of parameter values.
We may further expand the potential term in the limit ($m \ll g\phi^2$) as
\begin{equation}
(m+2g\phi^2)^{3/2}\approx (2g)^{3/2}\phi^3
\end{equation}
Substituting this expansion into the effective free-energy expression gives
\begin{equation}
F_{\mathrm{eff}}(\phi)=\frac{v}{24}\phi^4+\frac{w}{2}\phi^2-\frac{(2g)^{3/2}}{12\pi c^{3/2}}\phi^3 .
\end{equation}
The presence of the ($\phi^3$) term in the free-energy functional may induce a first-order phase transition in the system. Therefore, within this framework of competing magnetic phases, one may conclude that, in a geometrically frustrated system, although the phase is initially fluid-like, fluctuations and loop corrections can generate a stable minimum in the free-energy landscape of the competing phases. This provides the possibility of a phase transition even when such minima are absent in the original free-energy expression. This quantum phenomenon is similar to the Coleman-Weinberg\cite{ref-35} mechanism originally developed in the context of gravitational field models.\\

We further emphasize that, although different competing phases may coexist even in a pristine sample, the presence of inherent structural disorder can render the interaction spatially dependent. Around defects or sample termination sites, the competition between different phases can be locally altered. Consequently, fluctuation effects must be incorporated by considering a position-dependent interaction\cite{ref-31,ref-36,ref-46,ref-47}. Therefore, the fluctuation integral can be written as
\begin{equation}
\int \mathcal{D}[\zeta]\exp\left(-\frac{\beta}{2}\zeta M\zeta\right)
=(\det \beta M)^{-1/2},
\end{equation}
where
\begin{equation}
M=-c\nabla^2+m(\mathbf{r})+\frac{u}{2}\psi_c^2+2g\phi^2(\mathbf{r}).
\end{equation}
Since the coupling varies spatially, the fluctuation operator is no longer translationally invariant. A standard approach is to perform a derivative expansion\cite{ref-31,ref-36,ref-46,ref-47} of the functional determinant, retaining the local effective potential together with the leading-order gradient corrections. The principal consequence is that fluctuation corrections lead to a renormalization of the stiffness associated with the gradient term, thereby requiring the order parameter ($\phi(\mathbf{r})$) to be treated as a spatially varying field rather than a uniform one\cite{ref-31,ref-36,ref-46,ref-47}. Without presenting the full derivation, the resulting effective free-energy functional is given by\cite{ref-31,ref-36,ref-46,ref-47}:
\begin{equation}
F_{\mathrm{eff}}(\phi,\mathbf{r})=\int d^2r\left[(a_m-gX(\mathbf{r}))\phi^2(\mathbf{r})+c_m|\nabla\phi(\mathbf{r})|^2+b_m\phi^4(\mathbf{r})\right].
\end{equation}
where $a_m$, $c_m$, $b_m$ are coefficients, and ($X(\mathbf r)$) represents a random disorder field that locally modifies the quadratic coefficient of the order parameter. The corresponding partition function can then be expressed as
\begin{equation}
Z(X)=\int \mathcal{D}(\phi) \exp\left[-\beta\int d^2r\left\{(a_m-gX(\mathbf{r}))\phi^2(\mathbf{r}) + c_m|\nabla\phi(\mathbf{r})|^2+b_m\phi^4(\mathbf{r})\right\}\right].
\end{equation}
For simplicity, we assume ($X(\mathbf r)$) to be a Gaussian quenched disorder field with $\langle X(\mathbf r)\rangle=X_0$, and $\langle (X(\mathbf r)-X_0)(X(\mathbf r')-X_0)\rangle
=\Delta\delta(\mathbf r-\mathbf r')$.
Equivalently, the disorder correlation function can be written as
\begin{equation}
\langle X(\mathbf r)X(\mathbf r')\rangle
=X_0^2+\Delta\delta(\mathbf r-\mathbf r').
\end{equation}
The corresponding probability distribution is given by
\begin{equation}
P(X)\propto \exp\left[-\frac{1}{2\Delta}\int d^2r\left(X(\mathbf r)-X_0\right)^2\right].
\end{equation}
We employ the well-known phenomenological framework of the glass transition\cite{ref-19,ref-33,ref-37} and introduce the replica partition function in the following form:
\begin{equation}
\resizebox{\linewidth}{!}{%
$\left\{Z^n \right\}=\int \mathcal{D}X,P(X)\int \prod_{\alpha=1}^{n}\mathcal{D}(\phi_\alpha)\exp\left[
-\beta\int d^2r\sum_{\alpha=1}^{n}\left\{(a_m-gX(\mathbf r))\phi_\alpha^2(\mathbf r)
+c_m|\nabla\phi_\alpha(\mathbf r)|^2+b_m\phi_\alpha^4(\mathbf r)\right\}\right].$
}
\end{equation}
The disorder-averaged free energy is then obtained using the replica limit:
\begin{equation}
\overline{F}=-\frac{1}{\beta}\overline{\ln Z},
\end{equation}
where
\begin{equation}
\overline{\ln Z}=\lim_{n\rightarrow0}\frac{\overline{Z^n}-1}{n}.
\end{equation}
To perform the disorder averaging, we expand the disorder field as $X(\mathbf r) = X_0 + \delta X(\mathbf r)$ and separate the disorder-dependent contribution as
\begin{equation}
S_{\delta X}=\int d^2r\left[\frac{(X(\mathbf r)-X_0)^2}{2\Delta}-\beta g(X(\mathbf r)-X_0)\sum_{\alpha=1}^{n}\phi_\alpha^2(\mathbf r)\right].
\end{equation}
Integrating over the Gaussian disorder field yields an effective replicated partition function of the form
\begin{equation}
\int\prod_{\alpha=1}^{n}\mathcal{D}(\phi_\alpha)\exp(-S_{\mathrm{eff}}),
\end{equation}
where the effective action is given by
\begin{equation}
\sum_{\alpha=1}^{n}\beta\int d^2r\left[(a_m-gX_0)\phi_\alpha^2(\mathbf r)+c_m|\nabla\phi_\alpha(\mathbf r)|^2
+b_m\phi_\alpha^4(\mathbf r)\right]-\frac{\beta^2g^2\Delta}{2}\int d^2r\left(\sum_{\alpha=1}^{n}\phi_\alpha^2(\mathbf r)\right)^2 .
\end{equation}
We may expand the additional term generated by the disorder averaging as
\begin{equation}
\sum_{\alpha=1}^{n}\phi_\alpha^4(\mathbf r)+2\sum_{\alpha<\gamma}\phi_\alpha^2(\mathbf r)\phi_\gamma^2(\mathbf r).
\end{equation}
Substituting this expansion into the effective action gives
\begin{equation}
\resizebox{\linewidth}{!}
{$\sum_{\alpha=1}^{n}\int d^2r\left[\beta(a_m-gX_0)\phi_\alpha^2(\mathbf r)+\beta c_m|\nabla\phi_\alpha(\mathbf r)|^2+\left(\beta b_m-\frac{\beta^2g^2\Delta}{2}\right)
\phi_\alpha^4(\mathbf r)\right]-\beta^2g^2\Delta\int d^2r\sum_{\alpha<\gamma}\phi_\alpha^2(\mathbf r)\phi_\gamma^2(\mathbf r).$}
\end{equation}
We may now apply the Hubbard-Stratonovich transformation\cite{ref-19,ref-48} to decouple the inter-replica quartic interaction using the Gaussian identity
\begin{equation}
\int\mathcal{D}(Q)\exp\left[\int d^2r\left\{-\sum_{\alpha<\gamma}\frac{Q_{\alpha\gamma}^2(\mathbf r)}{4\beta^2g^2\Delta}
+\sum_{\alpha<\gamma}Q_{\alpha\gamma}(\mathbf r)\phi_\alpha(\mathbf r)\phi_\gamma(\mathbf r)\right\}\right].
\end{equation}
We can then obtain the saddle-point relation\cite{ref-34} by minimizing the resulting effective action with respect to the auxiliary field ($Q_{\alpha\gamma}$), following the standard procedure.
The saddle-point relation\cite{ref-34} can be written as
\begin{equation}
Q_{\alpha\gamma}=2\beta^{2}g^{2}\Delta\left\langle\phi_{\alpha}(\mathbf{r})\phi_{\gamma}(\mathbf{r})\right\rangle .
\end{equation}
Following the standard Sherrington--Kirkpatrick model\cite{ref-49} and the mean-field analysis, including the self-consistency equation method\cite{ref-49,ref-50,ref-51}, the instability condition near the glass transition temperature ($T=T_g$) can be written as
\begin{equation}
1=2\beta^{2}g^{2}\Delta.
\end{equation}
Therefore, the glass transition temperature is given by
\begin{equation}
T_g =\frac{\sqrt{2}\,g\sqrt{\Delta}}{k_B}.
\end{equation}
\begin{figure}[t]
\begin{center}
\includegraphics[width=1.0\columnwidth]{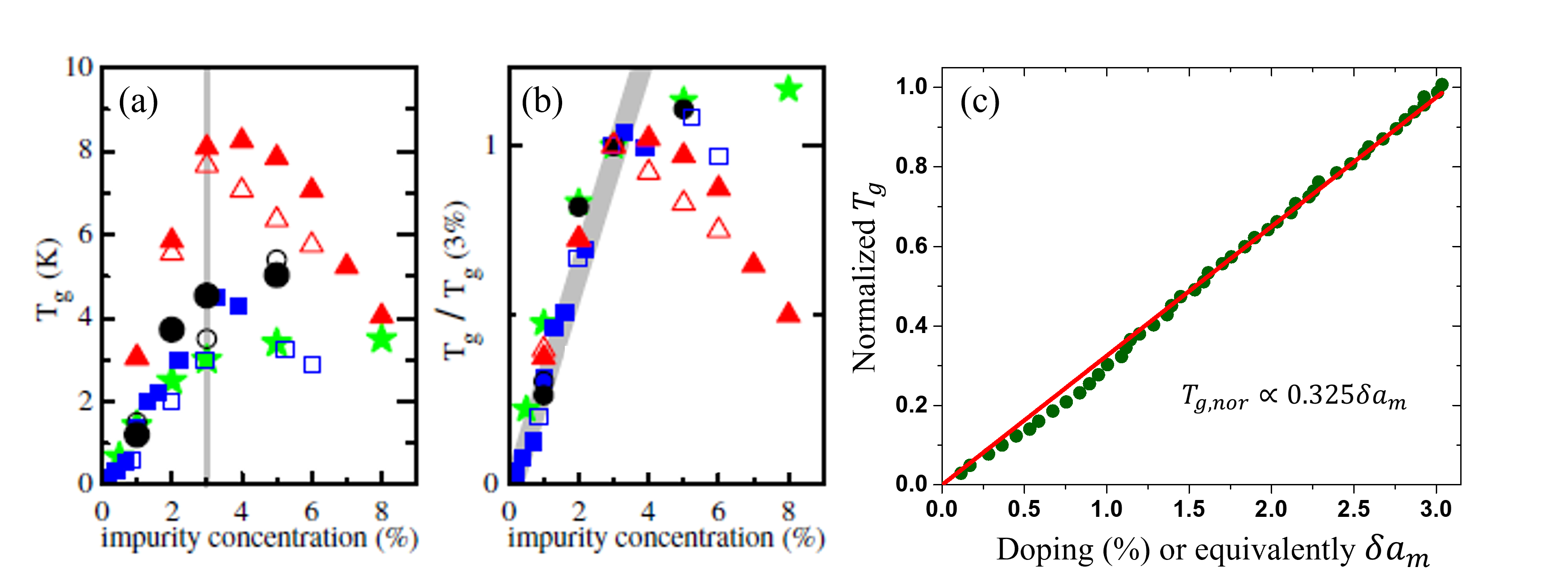}
\caption{Linear increase of the freezing temperature with increasing doping concentration in a selected low-dimensional material. (a) Variation of ($T_g$). (b) Variation of the normalized ($T_g$) (reproduced from \textbf{Reference 28} with permission). (c) Linear fit to the normalized ($T_g$) data (digitized from panel (b)) as a function of the doping concentration, which is treated as an effective measure of the disorder strength represented by the parameter ($\delta a_m$). In the low-doping limit, this allows us to approximately express the normalized glass transition temperature as $T_{g,\mathrm{nor}} \propto 0.325 \delta a_m.$}
\label{fig1}
\end{center}
\end{figure}
Therefore, within this simplified analogy, one expects a glass transition in the system. The transition temperature depends on the interaction strength and on the disorder variance parameter induced by the loop correction discussed earlier. Furthermore, $T_g$ is directly related to the amplitude of the random fluctuations, $\delta a_m \approx g\sqrt{\Delta}$. Within this simple phenomenological framework, the competition among different phases present in a geometrically frustrated antiferromagnetic system, whose relative strengths are modified by structural disorder, together with the additional effects of fluctuations, may give rise to either an order-by-disorder mechanism or glassy dynamics. Neither of these phenomena is captured by the original free-energy functional.
Interestingly, such a linear dependence of the glass transition temperature on the disorder strength, $T_g \propto \delta a_m$, has been reported for several low-dimensional materials\cite{ref-28}. In the low-doping limit, comparison of \textbf{Equation 33} with the available experimental and theoretical results\cite{ref-28} yields the approximate normalized glass transition temperature, $T_{g,\mathrm{nor}}\propto 0.325\,\delta a_m$, as illustrated in \textbf{Figure 2}.\\

In conclusion, we have demonstrated, within a simple phenomenological framework, that the competition among different phases in the fluid-like systems studied here, combined with fluctuations, can stabilize an ordered ground state even when the underlying free-energy functional does not intrinsically favor any ordered phase. Nevertheless, the unavoidable structural inhomogeneities and disorder suppress the development of true long-range order, resulting instead in the emergence of a glass-like state. The emergence of such glassy behavior may, under certain conditions, mimic spin-liquid-like behavior\cite{ref-6}, thereby making it difficult to identify genuine spin-liquid systems. These arguments are based on the Coleman-Weinberg mechanism\cite{ref-35}, which has been widely used to understand fluctuation-induced mass generation in a variety of gravitational phenomena.

\section*{Acknowledgments}

The author gratefully acknowledges the Indian Institute of Science for providing the facilities and infrastructure necessary to carry out this research. Artificial intelligence (AI) tools were used to assist with portions of the derivations and to improve the clarity of the manuscript. All AI-assisted content was carefully reviewed, verified, and critically evaluated by the author, who takes full responsibility for the accuracy, originality, and integrity of the final manuscript.



\end{document}